\documentclass[pdflatex,referee, sn-nature, Numbered]{sn-jnl}

\usepackage{graphicx}%
\usepackage{multirow}%
\usepackage{amsmath,amssymb,amsfonts}%
\usepackage{amsthm}%
\usepackage{mathrsfs}%
\usepackage[title]{appendix}%
\usepackage{xcolor}%
\usepackage{textcomp}%
\usepackage{manyfoot}%
\usepackage{booktabs}%
\usepackage{algorithm}%
\usepackage{algorithmicx}%
\usepackage{algpseudocode}%
\usepackage{listings}%
\usepackage{gensymb}
\usepackage{aas_macros}
\newcommand{\h}{\textsuperscript{h}}
\newcommand{\m}{\textsuperscript{m}}
\newcommand{\s}{\textsuperscript{s}}

\setcitestyle{super,sort&compress}
\newcommand{\frbname}{FRB 20191203A} 
\newcommand{\frbsnrmax}{10.7\,}

\newcommand{\frbra}{$05\h\,51\m\,58\s$\,} 
\newcommand{\frbraerr}{8s\,} 
\newcommand{\frbdec}{$-13^\circ\,18'$\,} 
\newcommand{\frbdecerr}{\ensuremath{4'}\,} 
\newcommand{\frbradecerr}{\ensuremath{4'}\,} 
\newcommand{\frbgallon}{218.45}
\newcommand{\frbgallat}{-19.1}
\newcommand{\frbfluence}{$360 \pm 30$\,Jy\,ms\,}
\newcommand{\frbpeakflux}{3.9\,Jy}
\newcommand{\frbtau}{$\tau=$90$\pm$20\,ms\,}

\newcommand{\frbdm}{393}
\newcommand{\frbdmmw}{66}
\newcommand{\frbdmeg}{277}
\newcommand{\frbdmhalo}{50}
\newcommand{\centralfreq}{154.88}

\newcommand{\dmeg}{\ensuremath{{\rm DM_{eg}}}} 
\newcommand{\dmmw}{\ensuremath{{\rm DM_{MW}}}} 
\newcommand{\dmtot}{\ensuremath{{\rm DM_{total}}}} 
\newcommand{\dmigm}{\ensuremath{{\rm DM_{IGM}}}} 
\newcommand{\dmhost}{\ensuremath{{\rm DM_{host}}}} 
\newcommand{\dmneb}{\ensuremath{{\rm DM_{nebula}}}} 
\newcommand{\dmu}{\ensuremath{{\rm pc\,cm^{-3}}}} 

\newcommand{\Tobs}{\ensuremath{T_{\rm obs}}}

\begin{document}
\title{A fast radio burst from a non-targeted low-frequency survey}


\author*[1,2]{\fnm{Cristian} \spfx{Di} \sur{Pietrantonio}}\email{cristian.dipietrantonio@csiro.au}
\author[3]{\fnm{Marcin} \sur{Sokolowski}}
\author[2]{\fnm{Clancy W.} \sur{James}}
\author[2]{\fnm{N. D. Ramesh} \sur{Bhat}}
\author[2,4]{\fnm{Randall} \sur{Wayth}}
\author[2,4]{\fnm{Danny C.} \sur{Price}}
\author[1]{\fnm{Christopher} \sur{Harris}}
\author[2]{\fnm{Chia Min} \sur{Tan}}
\author[2]{\fnm{Christopher P.} \sur{Lee}}
\author[5]{\fnm{Bradley W.} \sur{Meyers}}
\author[1]{\fnm{Mahitha} \sur{Ega}}
\author[4]{\fnm{Samuel J.} \sur{McSweeney}}
\author[1]{\fnm{Christopher} \sur{Schlipalius}}
\author[2]{\fnm{Gregory} \sur{Sleap}} 
\author[7]{\fnm{Steven E.} \sur{Tremblay}}
\author[6]{\fnm{Mengyao} \sur{Xue}} 
\affil*[1]{\orgname{Pawsey Supercomputing Research Centre},
    \orgaddress{Kensington, WA 6151, Australia}}
\affil[2]{International Centre for Radio Astronomy Research (ICRAR), Curtin University, Bentley, WA 6102, Australia}
\affil[3]{Australian SKA Regional Centre (AusSRC), University of Western Australia, 35 Stirling Highway, Crawley, WA 6009, Australia}
\affil[4]{SKA Observatory (SKAO), Kensington, WA 6151, Australia}
\affil[5]{Australian SKA Regional Centre (AusSRC), Curtin University, Kent Street, Bentley, WA 6102, Australia}
\affil[6]{National Astronomical Observatories, Chinese Academy of Sciences, 20A Datun Road, Chaoyang District, Beijing, China}
\affil[7]{National Radio Astronomy Observatory, 1011 Lopez Road, Socorro, NM 87801, USA}

\maketitle

\textbf{ 
Fast radio bursts (FRBs) are bright, millisecond-duration transient radio signals of extragalactic origin with unknown source and emission mechanisms \cite{Lorimer2007}. While several thousand FRBs have been discovered at frequencies above 400 MHz \cite{Amiri2021, Abbott2026}, detections at lower frequencies have been exceptionally rare \cite{Parent2020, Kumar2024}. Contributing to their rarity are the magnified dispersive delay, in the tens of seconds, and signal scattering that make low-frequency searches with established techniques computationally prohibitive. In particular, the only confirmed detection below 300\,MHz came from a Low-Frequency Array \cite{Haarlem2013} observing campaign targeted at FRB 20180916B \cite{PastorMarazuela2021,Pleunis2021,Gopinath2023}, a source already known to emit at higher frequencies. Here we present the first non-targeted detection of an FRB below 300\,MHz, achieved in a wide-field, imaging-based survey with the Murchison Widefield Array\cite{Tingay2013}. The implied FRB rate is consistent with that observed at higher frequencies, demonstrating that non-targeted wide-field surveys can uncover an unbiased, large low-frequency FRB population. This result challenges models invoking severe scattering or absorption to explain previous non-detections \cite{Sokolowski2018,Ravi2019,Chawla2025}, and strengthens the case of low-frequency observations as promising avenue to study the local environment of FRBs.
}
\\

Multiple unsuccessful attempts to discover FRBs at frequencies below 300\,MHz have been made in the last fifteen years with the Low-Frequency Array (LOFAR) \citep{Coenen2014, Karastergiou2015} and the Murchison Widefield Array\footnote{\url{https://www.mwatelescope.org}} (MWA) \citep{Tingay2015,  Rowlinson2016, Sokolowski2024, Kemp2024}. Because non-targeted low-frequency FRB searches are computationally challenging, observations targeting known repeating sources have been considered the only viable way to detect FRBs below 300\,MHz, as demonstrated by the unique detection of the repeating FRB 20180916B by LOFAR \cite {PastorMarazuela2021,Pleunis2021}. However, similar attempts targeting other repeating FRBs have not succeeded \cite{Tian2022,Chawla2025,McKenna2026}, and it has therefore remained unclear whether low-frequency emission is a property unique to that source.
Furthermore, repeating and non-repeating FRBs might still be two distinct populations\cite{Cook2026}. Hence, non-targeted searches are essential to retrieve an unbiased sample of low-frequency FRBs, necessary for population studies and cosmology applications. Finally, low-frequency detections can provide a wealth of information about FRB progenitors and their local environments because these are much more sensitive to propagation effects like scattering, dispersion, and Faraday rotation, and free-free and synchrotron self absorption, as they all scale with frequency $\nu$ as $\propto \nu^{\gamma}$, where $\gamma \le -2$ \citep{2012hpa..book.....L,1986rpa..book.....R}.

\begin{figure}
	\centering
	\includegraphics[width=\linewidth]{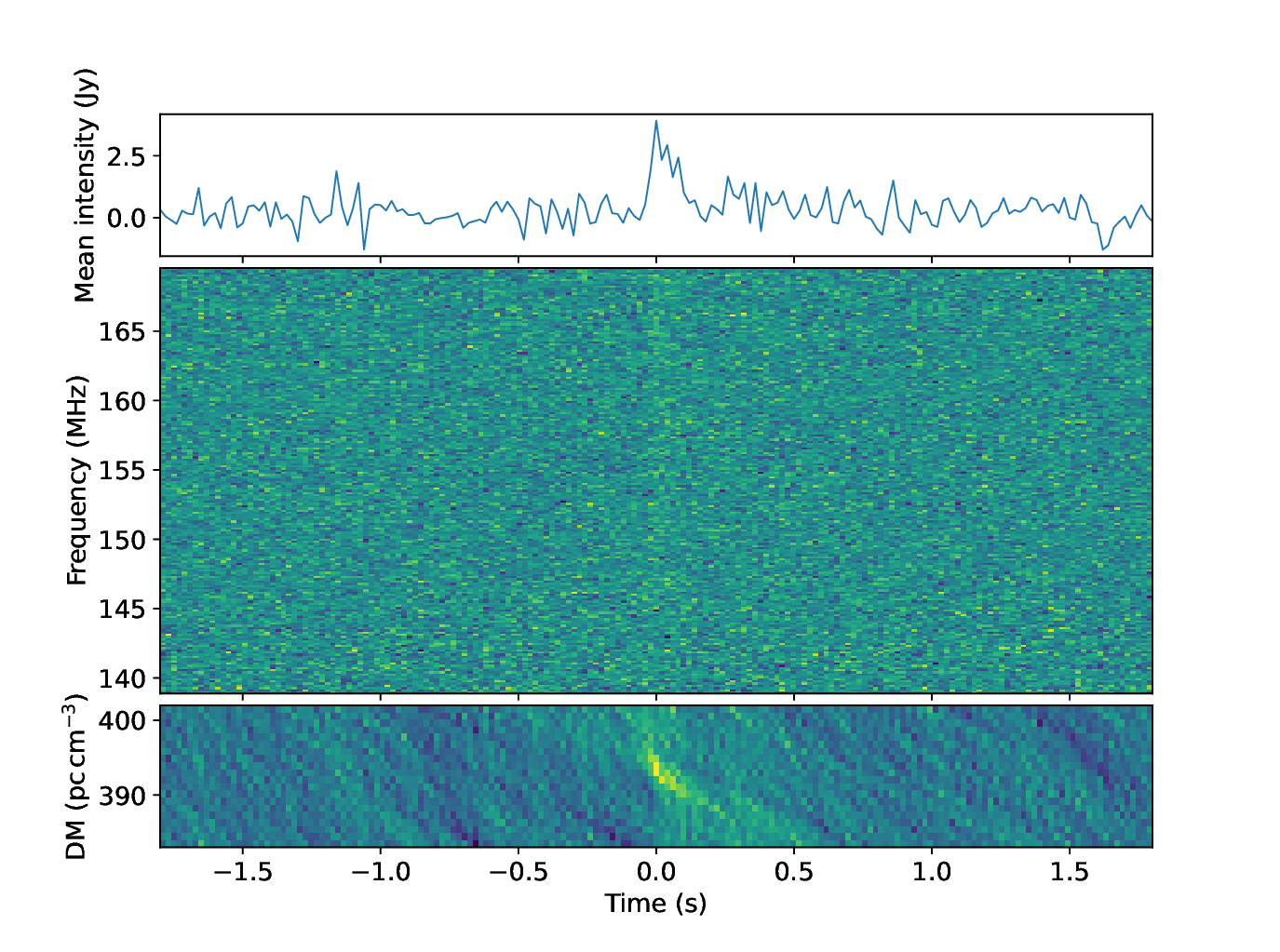} 
\caption{\textbf{Dedispersed time series and dynamic spectrum of \frbname{}.} Intensity values of image pixels corresponding to the FRB location are arranged into a dynamic spectrum at 40\,kHz and 20\,ms resolution. It covers the entire $30.72$\,MHz MWA band and a time span of $30$\,s, a few seconds more than the dispersive delay. The dynamic spectrum is then dedispersed at ${\rm DM} = \dmtot = 393{\rm \,pc\,cm^{-3}}$, contiguous channels are averaged to 160\,kHz, and the time span cropped down to 3 seconds around the time of arrival of the FRB. The result is presented in the centre panel. The top panel shows the dedispersed time series at 20\,ms time resolution, averaged across the band. The DM-time plot in the bottom panel shows the bow-tie structure typical of a dispersed astronomical radio signal.} 
\label{fig:detection}
\end{figure}

\begin{figure}
    \centering
    \includegraphics[width=\linewidth,trim={8em, 5em, 8em, 5em}]{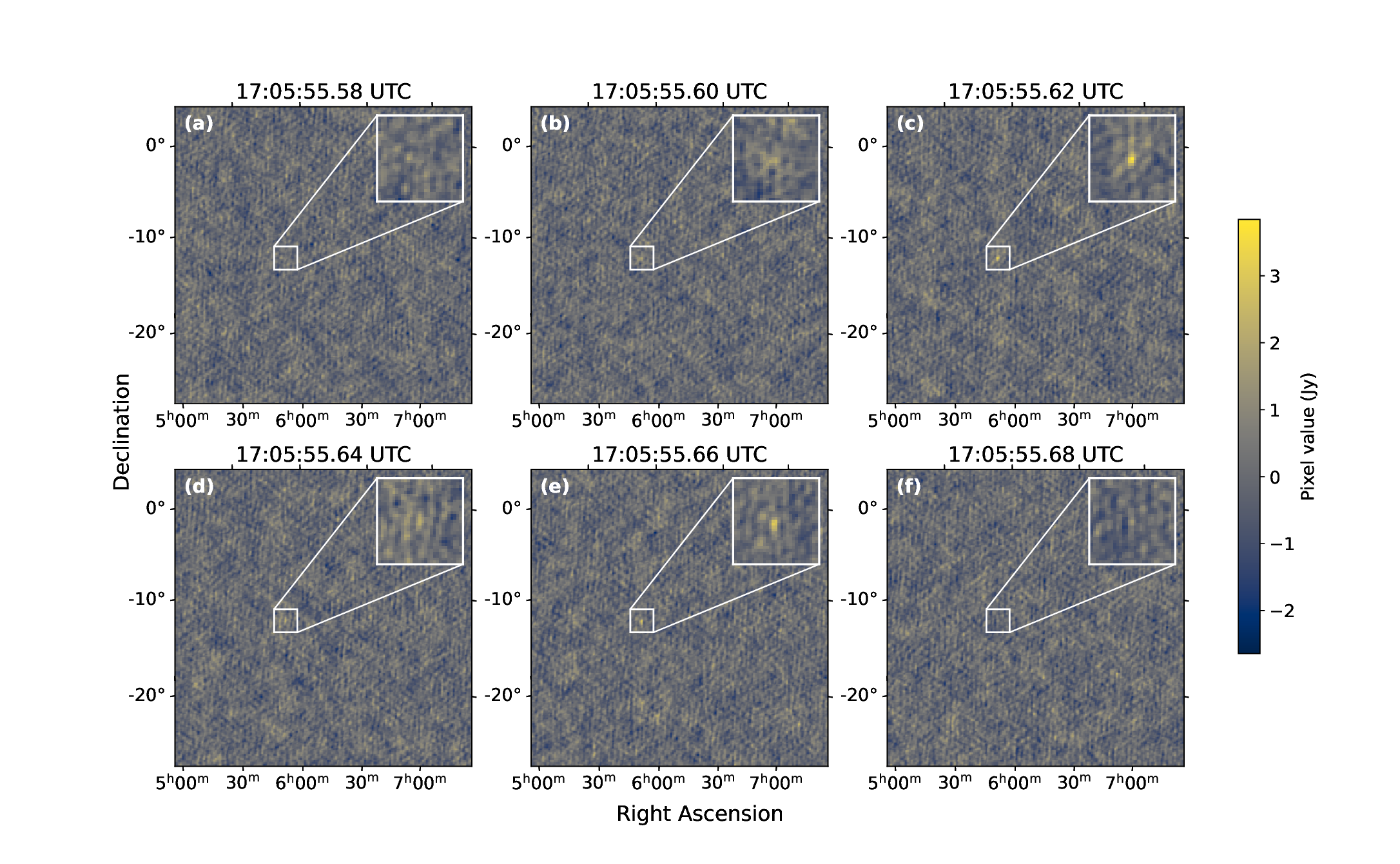}
    \caption{\textbf{Dedispersed difference images of \frbname{}}. Snapshot images covering the time extent of the dispersive delay are generated by the BLINK pipeline at 20\,ms and 40\,kHz resolutions, then dedispersed to the DM of the FRB, and averaged over frequency channels. The result is a time series of dedispersed images. The per-pixel median image is computed and then subtracted from the images before, during and after the peaks of \frbname{}. These are arranged chronologically from left to right, then top to bottom. An inset placed in the top-right corner zooms in on the FRB location. The FRB appears in images (b), (c) at its brightest, (d) and (e). The local maximum in (e) is brighter than the local minimum in (d) due to noise fluctuations. In the last image, (f), the signal disappears leaving only the noise.}
    \label{fig:image_panel}
\end{figure}

Here we present the first non-targeted detection of an FRB at frequencies below 300\,MHz. \frbname{}, shown in Fig. \ref{fig:detection}, was recorded on December 3, 2019 between frequencies 138.88 and 169.6\,MHz. The burst reached the top end of the frequency band at the telescope at 17:05:55.60 UTC and it is characterised by a signal-to-noise ratio (S/N) of \frbsnrmax{} when the time series is downsampled to the $120$\,ms width of the boxcar filter maximizing S/N. The detection occurred during an ongoing non-targeted, single-pulse search in the Southern-sky MWA Rapid Two-metre (SMART) survey for pulsars and fast transients at low frequencies \citep{Bhat2023}. The survey is a collection of 71, 80-minute voltage-capture observations that provide an effective exposure of $18,244$\,deg$^2$\,h, mainly due to the 610\,deg$^2$ field of view (FoV) of the MWA (see Methods). To process the nearly 4\,PB dataset from the SMART survey we developed a novel, high-time-resolution imaging pipeline leveraging graphics processing units (GPUs), called Breakthrough Low-latency Imaging with Next-generation Kernels (BLINK) \cite{GPUImager,DiPietrantonio2025}, and executed it on the Setonix supercomputer at the Pawsey Supercomputing Research Centre\footnote{\url{https://pawsey.org.au}}. The search used a 20\,ms time and 40\,kHz frequency resolution, and generated and analysed more than 80 billion images, for a total data volume of approximately 21\,PB, over the course of six months. Fig. \ref{fig:image_panel} displays the sequence of dedispersed 20\,ms difference images capturing the FRB in the instants before, during, and after its brightest peak. Extended Data Fig.~\ref{fig:dispersion_sweep} displays the dispersion sweep in the dynamic spectrum before dedispersion.

The measured fluence of $\mathcal{F} =$ \frbfluence places \frbname{} among the brightest FRBs ever detected \citep{Amiri2021,Abbott2026, 2025PASA...42...36S,Gopinath2023}. Its position is (RA, DEC) = (\frbra, \frbdec), which corresponds to Galactic coordinates $(l, b) = (\frbgallon^\circ, \frbgallat^\circ)$. The localisation error is about $\frbradecerr$. No robust host galaxy association is possible due to the insufficient spatial resolution of the MWA Phase II compact configuration (see Methods). 

Cold plasma delays the arrival time of a radio signal quadratically as it extends towards lower frequencies, and proportionally to the integrated electron column density along the propagation path, which is referred to as the dispersion measure (DM) \cite{Petroff2019}. \frbname{} is subject to a dispersive delay of 28.1\,s between the times of arrival at both ends of the 30.72\,MHz observing band, corresponding to a total DM, $\dmtot$, of $\frbdm\,\dmu$. A large DM value off the Galactic Plane is an indication that the signal has travelled through a significant amount of ionised gas that cannot be attributed solely to the interstellar medium. The latest electron density model NE2025 \cite{Ocker2026} predicts a Milky Way contribution $\dmmw$ of $\frbdmmw\,\dmu$ along the associated line of sight. We further attribute $\frbdmhalo\,\dmu$ to the Milky Way halo \citep{Cook2023} leading to the estimated extragalactic contribution $\dmeg$ of $\frbdmeg\,\dmu$. Assuming $\dmeg$ is entirely due to the intergalactic medium (IGM), we infer a maximum redshift $z \approx 0.3$ \citep{Macquart2020} and luminosity distance of approximately $1.55$\,Gpc. If this is the case, the band-limited isotropic energy of the burst is $E_{\text{iso}} \approx 8 \times 10^{41}$\,erg, which places the burst at the upper end of FRB energy distribution. The isotropic spectral luminosity calculated from the peak flux density of \frbpeakflux{} is $L\approx 10^{34}\,{\rm erg\,s^{-1}\,Hz^{-1}}$.

\begin{table*}[ht] 
    \centering
    
    \caption{\textbf{\frbname{} parameters.} The table summarises the most important properties of the detected FRB inferred from our data.}
    \label{tab:parameter_values}
    
    \begin{tabular}{@{} l c @{}} 
        \toprule
        \textbf{Parameter} & \textbf{Value} \\
        \midrule
        Peak flux density & $3.9 \pm 0.6$\,Jy \\
        Fluence & 360 $\pm$ 30\,Jy\,ms \\
        S/N & \frbsnrmax \\
        $\dmtot$ & \frbdm\,\dmu\\
        $\dmmw$ & \frbdmmw\,\dmu\\ 
        $\dmeg$ & \frbdmeg\,\dmu\\
        Equatorial coordinates (J2000)\textsuperscript{a} & \frbra, \frbdec \\ 
        Galactic coordinates & $\frbgallon^\circ, \frbgallat^\circ$ \\
        Scatter broadening time ($\tau$) & 90 $\pm$ 20\,ms \\
        \bottomrule
    \end{tabular}
    
    \vspace{4pt}
    \parbox{\linewidth}{\raggedright \footnotesize \textsuperscript{a} Position error is approximately $4'$.}

\end{table*}

Approximately 2.4\% of FRB sources have been observed to repeat\cite{Cook2026} and we investigated whether \frbname{} could be one of those. No additional burst was found in 4.5\,h of SMART and other MWA archival observations covering the FRB position. The Australian Square Kilometre Array Pathfinder (ASKAP), co-located with the MWA, was conducting a commensal FRB incoherent sum survey at frequencies between 0.7 and 1.8\,GHz \cite{2025PASA...42...36S} at the time of the burst. However, the FRB location was far from the beam centre, and no high-frequency counterpart was detected. Three hours of follow-up observations in the 0.7-4\,GHz frequency range with Murriyang, the 64-m telescope at Parkes, also resulted in a non-detection. The modest follow-up time and the combination of relatively low S/N and high fluence of the FRB prevent us from precluding the possibility of it being a repeater with typical bursts of lower fluences. For instance, the Canadian Hydrogen Intensity Mapping Experiment (CHIME) detected FRB 20121102A only once despite it being a known repeater \cite{Cook2026}.

Scatter broadening arises when turbulent, ionised gas in the intervening medium diffracts the astrophysical signal into multiple paths, producing delayed arrivals that broaden the pulse profile. This is observed as an exponential tail, $\propto \exp(-t/\tau)$, as a function of time, $t$, and $\tau$ is the scatter broadening time. Assuming a thin screen model \citep{2003ApJ...584..782B}, a Gaussian function convolved with an exponential tail\citep{2019ApJ...874..179K} was fitted to the 20\,ms time series (Extended Data Fig. ~\ref{fig:scatter_broadening_time_fit}). The fit resulted in \frbtau, and an upper limit on the intrinsic pulse width of less than 20\,ms.

Scattering scales with frequency as $\tau(\nu) \propto \nu^{\beta}$, where $\beta$ is the scattering index. For pure Kolomogorov turbulence, $\beta = -4.4$, with similar values of $\beta \approx -4$ observed in a large sample of pulsars \citep{2004ApJ...605..759B}. However, much shallower values ranging down to even $\beta=-1$ have been observed in some pulsars \citep{2017MNRAS.470.2659G,2017ApJ...846..104K} and FRBs \citep{2025PASA...42..133S}. While we cannot constrain the frequency dependence of scattering in \frbname{} due to its low S/N and the narrow observing bandwidth, scatter broadening time is expected to increase at low frequencies in all scenarios. 

\begin{figure}
    \centering

    \includegraphics[width=\linewidth]{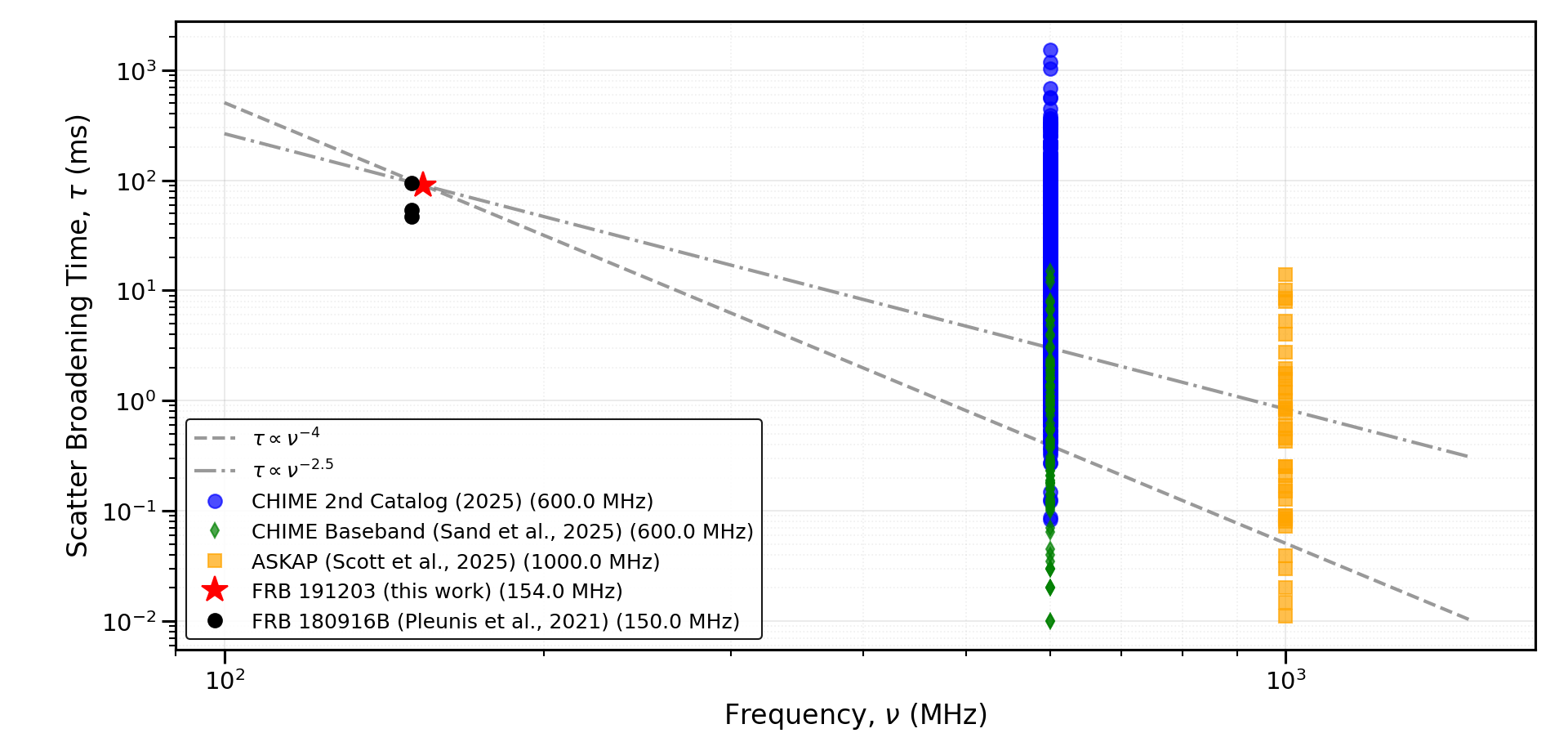}
    \caption{\textbf{Scatter broadening time ($\tau$) of FRBs as a function of frequency.} The red star data point is the $\tau$ of \frbname{} measured in this work at \centralfreq \,MHz. For reference, black data points are scatter broadening times of 3 bursts from FRB 190816B detected by LOFAR at 150\,MHz \citep{Pleunis2021}, blue points are 1193 CHIME FRBs from the second catalogue \citep{Abbott2026}, green points represent 137 CHIME FRBs in the baseband catalogue \citep{2025ApJ...979..160S}, and yellow points are 35 ASKAP FRBs from high-time resolution measurements \citep{2025PASA...42..133S}. Upper limits are not included. The dashed and dashed-dotted lines corresponds to scatter broadening time of \frbname{} scaled to other frequencies with scattering index -4 and -2.5 respectively. For the scattering index -4 (dashed line), it was calculated that 49 out of 137 CHIME baseband FRBs, 5 out of 35 ASKAP FRBs, and 12 out of 1993 FRBs from the CHIME second catalogue are less scattered than \frbname{}, placing our detection among the less scattered FRBs.}
    \label{fig:scatter_broadening_vs_freq}
\end{figure}

LOFAR detections of FRB 20180916B \citep{Pleunis2021,Gopinath2023} presented scattering timescales between 40 to 60\,ms, scaled to \centralfreq\,MHz, slightly less than the measured value of \frbname{}. To compare with a larger FRB population, we used scatter broadening time measurements of 137 CHIME FRBs in baseband data at 600\,MHz \citep{2025ApJ...979..160S} and 35 ASKAP FRBs observed in high-time resolution data at 1000\,MHz \citep{2025PASA...42..133S}. We scaled our $\tau$ measurement at \centralfreq\,MHz to other frequencies, using scattering indices -2.5, the mean scattering index in the ASKAP sample, and -4 \citep{2003ApJ...584..782B}. The comparison is shown in Fig. ~\ref{fig:scatter_broadening_vs_freq}. The scatter broadening time of \frbname{} scaled to 1000\,MHz places it among the 14\% of the least scattered ASKAP FRBs, while $\tau$ scaled to 600\,MHz falls closer to the middle of the CHIME baseband sample: 49 out of 137 FRBs are less scattered. Overall, scatter broadening time of \frbname{} is consistent with higher frequency measurements, albeit in the less scattered part of the population. Most importantly, this detection confirms that FRBs can be detected below 200\,MHz despite the larger degree of scattering at low frequencies.

Given the FRB detection and the effective exposure of our search, we infer an FRB rate of 54\,sky$^{-1}$\,day$^{-1}$, with a 95\% confidence interval of 1-302\,sky$^{-1}$\,day$^{-1}$, for bursts with fluence above $57$\,Jy\,ms. This result is consistent with the rate of 3-450 FRBs\,sky$^{-1}$\,day$^{-1}$ above a fluence limit of 50\,Jy\,ms set by the LOFAR detection of FRB 20180916B \cite{PastorMarazuela2021}. Moreover, the FRB rate at low frequencies we infer is consistent with that of FRBs at higher frequencies, at the same fluence threshold, and assuming a Euclidean rate scaling and no intrinsic frequency dependence \citep{Sokolowski2024}. Contrary to what is proposed in the literature\cite{Sokolowski2018,Ravi2019,Chawla2025}, no emission suppression or rate turnover is observed. This implies FRBs are as common at low frequencies as they are at the frequency bands of CHIME and ASKAP.

Our result suggests that the lack of detections in past non-targeted searches can be attributed to insufficient exposure, sensitivity, or both. Previous searches \citep{Tingay2015, Rowlinson2016, Sokolowski2018, Kemp2024} had limited time and frequency resolutions on the order of seconds and MHz scales, respectively, which are too coarse over the 70-300\,MHz frequency range observed by the MWA. Dispersive delay and scattering temporally broaden an FRB signal reducing its detectability, since the FRB fluence gets diluted with noise. These propagation effects are more pronounced at low frequencies, leading to degradation of sensitivity and an increase of computational cost to recover the signal. Our software, BLINK, leverages novel algorithms \cite{DiPietrantonio2026}, GPU-accelerated software, and supercomputing infrastructure to process PB-scale data volumes, to take advantage of the full FoV of the MWA and its many hours of archival data, drastically increasing the exposure of the search, and to achieve the millisecond-time scale and kHz-resolution necessary to detect FRBs at low frequencies. 

Several hypotheses were advanced to explain non-detections in targeted searches, such as chromatic activity \citep{Houben2019, Tian2022} and absorption due to the local environment \citep{Sokolowski2018}. A recent large-scale search with LOFAR targeting 487 sources for a total of 725 hours, including 33 CHIME repeating FRBs, resulted in no detections \citep{Chawla2025}. This result does not contradict our rate estimate as only 2.4\% of sources are observed to repeat\cite{Cook2026}. Our detection demonstrates that non-targeted searches with wide-field, low-frequency interferometers are necessary to detect FRBs without prior knowledge of their location, thereby eliminating a major observational bias in studies of the low-frequency FRB population.

 Observations at low frequencies provide an effective probe of FRB local environments and constrain the size, age, and DM contribution of a surrounding nebula \citep{2016ApJ...824L..32P}. Despite predictions of strong absorption at low frequencies, the detection of \frbname{} and the earlier LOFAR detections of FRB 20180916B \cite{PastorMarazuela2021,Pleunis2021} show that at least in some cases the surroundings of FRBs are transparent to low-frequency radio waves. Most current models attribute FRBs to highly magnetised neutron stars (magnetars) \citep{2019MNRAS.485.4091M}, which can be formed in cataclysmic processes such as a supernova (SN) explosion. Analysis of bursts from actively repeating FRBs \cite{2026ApJS..284...77W}, and observations of persistent radio sources at their locations\citep{2023ApJ...958..185C}, are consistent with their progenitors being embedded in a young supernova remnant (SNR) or pulsar wind nebula \citep{2021ApJ...908L..10H,2025arXiv251207140W,2026SciBu..71...76N,2026ApJ..1000L..53P}. Assuming this is also the case for \frbname{} and using methods described in literature \cite{Pleunis2021,Sokolowski2018}, the size of the absorbing region and the age of the SNR can be constrained. If the entire extragalactic $\dmeg \approx \frbdmeg(1+z)\,\dmu$ is attributed to the expanding SNR, the size of the absorbing region due to this nebula can be limited to $\Delta L\gtrsim 2 (T/10^4$\,{\rm K})$^{-1.35}$\,pc. This can be translated to relatively broad limits on the age of the SNR: $\gtrsim$450\,years for the mass of ejecta of 10\,$M_{\odot}$, or $\gtrsim$180\,years for 1\,$M_{\odot}$ \citep{2016ApJ...824L..32P,1995PhR...256..157M}. More robust estimates on the DM contribution of the local FRB environment, $\dmneb$, are required to constrain the size and age of a nebula. For example, a 
 $\dmneb\sim 70(1+z)\,\dmu$, assumed in a previous work \cite{2020Natur.577..190M}, would limit the absorption region to smaller sizes $\gtrsim 0.13 (T/10^4$\,{\rm K})$^{-1.35}$\,pc and the age to $\gtrsim$10\,years. A secure identification of the FRB host galaxy would provide more stringent limits on $\dmneb$, by separating extragalactic DM into $\dmhost$ and $\dmigm$. Furthermore, accurate measurements of $\dmneb$ and rotation measure (RM) enable estimates of electron density and ejecta mass of the FRB local environments \cite{2018ApJ...861..150P}, but this requires arcsecond localisation of the FRB signal. 

 Future surveys with low-frequency telescopes such as MWA Phase III \cite{Tingay2026} and SKA-Low\footnote{\url{https://www.skatelescope.org}} \cite{Dewdney2009} will provide the angular resolution required to identify host galaxies of FRBs, and to test FRB formation scenarios which predict that FRBs in a post-SNR nebula will be absorbed for a few hundred years longer than GHz-frequency radio waves  \citep{2016ApJ...824L..32P}. Additionally, simultaneous multi-frequency searches and monitoring with high- and low-frequency telescopes \cite{Sokolowski2018, PastorMarazuela2021} will enable the study of their evolution and progenitors. With SKA-Low expected to detect hundreds to thousands of FRBs per year \citep{2022aapr.confE...1S,2026arXiv260627714C}, surveys at low frequencies will not only complement FRB exploration but will also uniquely contribute to advancing astrophysics with these cosmological probes.

\section*{Methods}

\subsection*{Telescope}
\label{sec:telescope}
The Murchison Widefield Array (MWA) is a low frequency, interferometric radio telescope located in a radio-quiet zone of Western Australia. It consists of 256 aperture arrays, also known as tiles, providing a FoV of $610$\,deg$^2$ at 150\,MHz. It observes at frequencies between 70 and 300\,MHz, with an instantaneous bandwidth of 30.72\,MHz partitioned in 24 coarse channels. The telescope was deployed in three phases. Phase I comprised 128 tiles of 16 dipole antennas and a maximum baseline of 3\,km \citep{Tingay2013}. Voltage data was captured at the tile level in coarse channels, that were then fine-channelised on-site with Field-Programmable Gate Arrays (FPGAs). Data could be then downsampled and correlated, or sent to a Voltage Capture System (VCS)\cite{Tremblay2015} to be archived at the native time and frequency resolutions of 100\,$\mu$s time and 10\,kHz. Phase II \citep{Wayth2018}, which is the one used to record the observations of the SMART survey used in this work, increased the maximum baseline to 5\,km and doubled the number of tiles installed on site, with up to 128 usable at any given time. The recent Phase III upgrade \citep{Tingay2026} enables the use of the full array with additional receivers and a new correlation system \citep{Morrison2023}, doubling the sensitivity of the instrument compared to Phase I.

\subsection*{The SMART survey}
\label{sec:smart}
The Southern-sky MWA Rapid Two-metre (SMART) survey \citep{Bhat2023} leverages the 610\,deg$^2$ FoV of the MWA and the VCS to undertake a search for pulsars and fast transients in the 140-170\,MHz band and across the entire sky below declination $+30^\circ$. It is the first all-sky pulsar survey in the Southern Hemisphere at frequencies also to be observed by SKA-Low \cite{Dewdney2009}. The survey was conducted in the compact configuration of the MWA Phase II. Results from a partially-completed shallow processing project were promising and included several new pulsar discoveries after only $\sim$5\% of the survey data were analysed, and re-detections of 120 known pulsars from initial data quality checks and analysis \citep{Bhat2023a}.
In the latest SMART data releases \citep{2025PASA...42..117L,Bhat2026}, a total of 245 known pulsars have been detected and characterised, with many new pulsar discoveries to be published in an upcoming paper.
SMART data have also been used to confirm pulsar candidates identified in image-based transient searches \citep{2025ApJ...981..143M}. 

The SMART survey observing campaign was completed in late 2023, collecting 71, 80-minute (4,800\,s) long observations, for a total data volume of approximately 4\,PB. It is the largest all-sky survey in the southern sky in terms of data volume, and the second largest, globally, after the LOFAR Tied-Array All-Sky (LOTAAS) survey \citep{lotaas}. Due to the significant data rate and volume, VCS data are challenging to store and process, especially for routine pulsar detection and analysis. However, archival voltage data can be reprocessed as better algorithms and faster software pipelines become available. The BLINK project is one such example where SMART voltage data is searched for fast transients using an innovative imaging-based approach.

\subsection*{Search method}
SMART observations were imaged at 20\,ms cadence and 40\,kHz frequency resolution using the purpose-built GPU-based BLINK pipeline \citep{GPUImager,DiPietrantonio2025}. The image size of $256^2$ pixels, each covering approximately $0.014$\,deg$^2$, encompasses the entire FoV of the instrument. Image noise, defined as the Root Mean Square (RMS) of pixel values in the central region of an image, is adopted as a quality metric. For every set of 1,600 images produced for a second of observation and a coarse channel, pixel values in images where the noise is three times the median are discarded and replaced with pixels sampled from good images. Empirically, we found the approach reduced the number of candidates per observation from up to several millions, caused by strong terrestrial radio-frequency interference (RFI), to the order of a hundred thousand.

The BLINK imaging pipeline was extended with the novel Streaming high-Time Resolution Image DEdispersion\cite{DiPietrantonio2026} (STRIDE) algorithm to incrementally compute per-pixel dedispersed time series as images are produced.  The trial DMs varied from $10\,\dmu$ to $600\,\dmu$ in steps of 1\,\dmu. Hence, Galactic transient sources such as pulsars and Rotating Radio Transients (RRATs) were also targeted at DMs less than 100\,\dmu, which added a relatively low computational cost compared to staging terabytes-sized voltage data into GPU memory and searching at higher DMs. Dedispersed time series were generated and analysed in 2-second time intervals, with the standard deviation of noise $\sigma_{\text{2s}}$ estimated using robust statistics through median and inter-quartile range.

The BLINK pipeline used a multi-level candidate selection process for each observation. At first, candidate pulses with $\text{S/N}_{{\text{2s}}}$ $\ge 8.5$ were identified in the 20-ms time series, where $\text{S/N}_{{\text{2s}}}$ is the signal-to-noise ratio calculated with $\sigma_{\text{2s}}$. This resulted in $\sim$10$^3$-$10^4$ raw candidates. Those occurring when the standard deviation of total power across the band exceeded $5\sigma$ above the median standard deviation, and the ones detected at positions of known pulsars, were excluded. Dynamic spectra were generated from the brightest $\sim$1000 candidates, covering a time span 5\,s longer than the dispersion delay of each candidate's DM. Experimentally, we found that $\sigma_{\text{2s}}$ could under- or overestimate the actual noise $\sigma$, and therefore the S/N, by up to 30\%, as $\sigma$ converges to a stable plateau on time series of approximately 20\,s in length. Hence, the S/N of the candidates was re-calculated using the entire extent of the dynamic spectra, in the tens of seconds, and a threshold S/N $\geq 5$ was applied. At this stage, $\sim$100 candidates remained for visual inspection, with the number of candidates reduced mainly by the total power anti-RFI criteria. This relatively complex process was established to make inspection of the best candidates feasible and practical. The number of candidates inspected corresponds to an effective S/N threshold of approximately 6.5, for which pure Gaussian noise data would produce $\sim$10 false positives from a 80-minute observation.

\subsection*{Compute environment}
The processing is still in progress at the time of writing, and it is performed on the Setonix supercomputer, hosted at the Pawsey Supercomputing Research Centre (Pawsey) in Perth, Western Australia. Setonix's GPU partition comprises 192 GPU nodes, each equipped with 4 AMD MI250X GPUs, one AMD 64-core Zen3 ``Milan'' CPU, and 256\,GB of host memory. A single AMD MI250X GPU consists of 2 Graphics Compute Dies, each featuring 7040 cores and 64\,GB of global memory. Effectively, each compute node presents 8 logical GPUs to the software. 

Pawsey is also home to the MWA data archive containing the observations of interest. The archive resides on a hybrid system composed of a tape library, known as Banksia, and a disk-based object storage, named Acacia. Observations are staged on the supercomputer's filesystem using the MWA node of the All-Sky Virtual Observatory (ASVO)\citep{2020PASA...37...21S}, a web application serving MWA data products stored on Pawsey infrastructure to the astronomy community. For this search, data is delivered on the SSD-based, 3\,PB-sized pool of the \texttt{/scratch} filesystem to minimise read latency during processing.

A SMART observation cannot be processed on a single run because the runtime would exceed the wall-time limit of 24\,h on Setonix. Hence, the search is partitioned along two dimensions: observation time, using 9 overlapping 10-minute segments, and DM, dividing the 590 DM trials in 7 disjoint intervals. Specifically, DMs up to and excluding 500\,$\dmu$ are processed in groups of 90 or 100 trials; the more computationally challenging DM range 500 to 600\,$\dmu$ is split in half. This approach results in 63 instances of the BLINK pipeline being submitted as jobs to the Setonix scheduler for every observation. Wall times vary from 8\,h to 19\,h, depending largely on the subset of DMs being explored, but also affected by the performance of the filesystem at the time of execution.

Each observation is processed in one to three days, according to the availability of compute resources, at an average compute cost of 600 GPU node-hours. The amount is equivalent to using the entire Setonix GPU partition for three hours. By imaging at 20\,ms and 40\,kHz resolution, each second of observation results in $38,400$ images covering $50$ time bins and $768$ channels. The same voltage data are imaged seven times to process all DM trial intervals. Hence, a total of $38,400 \times 4,800 \times 7 \approx 1.3$ billion images per observation are generated. After dedispersion, nine trillion ($\sim 10^{13}$) time series points are evaluated during the peak detection stage. Currently 80\% complete, the processing of the survey has consumed 30.3\,MWh of energy and close to 36,000 GPU node-hours.

\subsection*{\frbname{} verification}
The first indication that the signal is a genuine FRB was its detection in multiple neighbouring pixels, at similar DMs, from 392 to 394\,$\dmu$, and at two consecutive time bins, with S/N$_{{\text{2s}}}$ between 8 and 11. This is the typical pattern observed with pulsar and RRAT detections in this same search. Three raw candidates caused by the FRB made it to the visual inspection stage, among 137 other candidates, starting from the initial 155,000 generated for the observation. Furthermore, the FRB underwent an additional examination process designed to verify its  astrophysical nature, to confirm its coordinates, and to exclude any instrumental effects such as aliasing and sidelobe detections. In practice, this meant running the BLINK pipeline multiple times, exploring various combinations of image and pixel sizes, as well as re-phasing visibilities to place the candidate at the centre. The signal was re-detected on each occasion. Multiple known pulsars were detected during the search and only single pulses of astrophysical nature passed all the checks. The scattering tail highlighted in Extended Data Fig.~\ref{fig:scatter_broadening_time_fit} provided additional confirmation of the astrophysical nature of the signal. The detection was independently reproduced with VCSBeam, the beamforming software purposely built and used to survey the SMART dataset for pulsars \citep{ Bhat2026, 2022PASA...39...20S, Ord2019}.

We also wish to highlight that the Inyarrimanha Ilgari Bundara, the CSIRO Murchison Radio-astronomy Observatory where the MWA resides, is a pristine environment minimally impacted by significant RFI\cite{Offringa2015}, especially when compared to other observatories. This allows us to easily filter out artificial transmissions and to be sensitive to the faintest extragalactic signals. Finally, imaging removes self-correlations, reducing sensitivity to uncorrelated RFI signals, and it associates a point-like source to a far-field, main-beam event.

\subsection*{Flux density calibration}
\label{sec:flux_calibration}

The BLINK FRB search pipeline has primarily been designed as a highly efficient discovery tool. Hence, the precise flux density calibration has been performed in the off-line analysis. 

The BLINK pipeline uses calibration solutions from the MWA ASVO database, which are readily available. These were originally derived from observations of MWA primary calibrator sources near their transit. Hence, the flux density scale is, to within a scaling factor of $\sim$1, correct. Nevertheless, in order to further improve the accuracy of flux density calibration, correct for the slightly off-the-beam-centre position of the calibrator and other effects, the following multiplicative correction factors were applied to the BLINK dirty images:

\begin{enumerate}
\item To account for the calibrator source Hydra A being offset from the beam centre, the BLINK dirty images were multiplied by a correction factor of 0.94.
\item Primary beam correction at the position of \frbname{} leads to multiplication by 1.05, as it was detected very close to the beam centre.
\item Normalisation correction of dirty BLINK images to match the flux scale of CLEAN\citep{1974A&AS...15..417H} and dirty WSCLEAN\citep{2014MNRAS.444..606O} images ($\times$1.19) as these were used as a flux density reference. 
\end{enumerate}

Combining the above resulted in the final scaling factor of approximately 1.17. As an additional verification, this calibration factor was estimated as a ratio $g = \sigma_{\rm sim} / \sigma$, where $\sigma_{\rm sim}$ is the standard deviation of the noise simulated with the MWA 2016 Beam model \citep{2017PASA...34...62S}\footnote{\url{https://github.com/marcinsokolowski/mwa_pb_msok/}}, and $\sigma$ is the standard deviation of the time series noise (excluding the FRB). The resulting $g \approx 1.5$ provided us with an estimate of the systematic error $\sim$30\%.

As a final verification, and after the scaling factor of 1.17 was applied, the calibrated flux densities of the sources in the BLINK dirty image were compared with their flux density in the GaLactic and Extragalactic All-sky MWA (GLEAM) catalogue \citep{2017MNRAS.464.1146H,2015PASA...32...25W}. In the MWA compact configuration the Point Spread Function (PSF; i.e., the shape of point sources) is very complex and reflects the hexagonal configuration of the MWA array. For this reason, standard source finders typically used for MWA data, like AEGEAN \citep{2018PASA...35...11H}, cannot be applied to dirty or CLEAN images. Therefore, the continuum sources were first identified in the 30-second long-integration WSCLEAN CLEAN image (10000 minor cycle iterations) formed with the Image Domain Gridding \cite{2018A&A...616A..27V} algorithm. Then, flux densities of the sources identified in the CLEAN image were extracted from the long-integration BLINK dirty image and compared against the GLEAM catalogue. 

For about 30 sources identified in the WSCLEAN CLEAN image, flux densities were extracted from both the CLEAN WSCLEAN and BLINK dirty images and compared with their flux densities in the GLEAM catalogue at \centralfreq\,MHz. Due to the relatively large size of the synthesised beam in the MWA compact configuration ($\sim0.16^\circ$) nearby sources can be blended into a single source. Therefore, the flux densities of the BLINK sources were compared with the integrated flux density of all GLEAM sources within a matching radius of $13'$. Extended Data Fig. \ref{fig:blink_vs_gleam_flux} shows the flux densities extracted from the BLINK dirty image versus the integrated GLEAM flux density. The slope of the fitted curve (0.84 $\pm$ 0.12) is within the error consistent with the one-to-one line, which confirms the precision of the BLINK flux scale after the above corrections.

The calibrated lightcurve in 20\,ms time resolution with the fitted model is shown in Extended Data Fig.~\ref{fig:scatter_broadening_time_fit}.
The resulting peak flux density and fluence of the FRB are also in agreement, to within a factor of $\sim$2, with the values obtained from the calibration of beamformed time series used in pulsar searches of the SMART data \cite{2025PASA...42..117L, 2017ApJ...851...20M}.

The presented method demonstrates that the BLINK pipeline and the auxiliary post-processing tools are entirely self-sufficient and enable discovery and precise flux calibration of pulses from FRBs, pulsars and other fast transients using image-based methods.

\subsection*{Astrometric calibration}
\label{sec:astrometry}

The BLINK pipeline forms dirty images in the standard orthographic projection described by a set of World Coordinate System FITS keywords \citep{2002A&A...395.1077C}, which are correctly interpreted by standard FITS viewers like DS9 or CARTA \citep{2026PASP..138b4506W}. Therefore, the \frbname{} position RA = \frbra $\pm$ \frbraerr, DEC = \frbdec $\pm$ \frbdecerr was obtained from the difference image (Fig.~\ref{fig:image_panel}) by fitting the 2D Gaussian profile at the burst position using CARTA. The errors were estimated by performing the same fit to 40 sources in the 30\,s BLINK dirty image used for verification of flux density calibration. The fitted positions were compared with the GLEAM catalogue positions. Offsets are consistent with zero, while the standard deviations of the RA and DEC offsets from the GLEAM positions were adopted as the position errors. This uncertainty is dominated by the arcmin-level error due to the spatial resolution of the MWA compact configuration, while the arcsecond-level uncertainties in GLEAM are negligible. The larger standard deviation ($\frbradecerr$) of the resulting offsets in RA and DEC was adopted as the localisation uncertainty. The radius of the corresponding error circle at the 99.7\% confidence level is approximately $13'$.

\subsection*{Attempts to identify the host galaxy}
\label{sec:host_galaxy}

We attempted to identify potential host galaxies of \frbname{} in the WISE $\times$ SuperCOSMOS Photometric Redshift Catalog \citep{2016ApJS..225....5B} within the $13'$ (3$\sigma$) error box. However, the FRB position is on the edge of the region excluded from these optical catalogues due to dust near the Galactic Plane. Therefore, valid candidates are only present west of the FRB position the closest being at an angular distance of about $7'$, and there are no candidates to the east. Searches in other catalogues did not yield robust matches due to the combination of their limited sky, redshift coverage, and large FRB position uncertainty. We conclude that it is not possible to identify a statistically robust sample of candidate host galaxies within the entire error box.

\subsection*{FRB rate estimate}\label{sec:frb_rate}

We investigated the expected number of daily FRB events at 150 MHz across the whole sky\cite{Keane2015}. The rate calculation was based on the effective exposure of the experiment to FRBs, measured in deg$^2$\,h. The exposure quantifies the on-sky time and the FoV, and it only includes observations where FRBs can be detected. The SMART survey comprises 71 observations. We excluded from the exposure calculation 20 observations pointing towards or close to the Galactic Centre, where the dense interstellar medium reduces the chance of detecting extragalactic sources. These were still processed to remain open to unexpected discoveries. Additionally, 12 eligible observations have not been processed yet. Another one was excluded due to tens of thousands of candidates caused by the Crab pulsar. Next, the noise level of the remaining 38 observations was evaluated. BLINK stores the dynamic spectrum of the central pixel of each observation it processes. As a quality metric, the RMS of the power across that spectrum was computed for each time bin. We did not expect to detect FRBs in observations where the RMS is consistently high or presents extreme variations. We empirically found that in good quality observations RMS would not exceed 20 Jy, also validated both by the lack of RFI and the detection in such observations of several pulsars not known to have bright pulses. Hence, observations with RMS exceeding the threshold of 20 Jy were excluded, leaving 30 good quality observations. Further, 9 observations had a significant number of images flagged: 3.7\%, 5\%, 7\%, 7.1\%, 11.5\%, 13\%, 14\%, 19\%, and 79\% of the total, for which the exposure time was reduced accordingly. Finally, the last 40 seconds of each observation were flagged due to incomplete dispersive sweeps. The total exposure time $\Tobs$ is 148,932\,s or 41.37\,h.

The area of the sky covered is 441\,deg$^2$, computed as the median full-width, half-maximum of all beams across the selected 30 observations. Hence, the effective exposure is $E = {\rm FoV} \times \Tobs = 441\,\text{deg}^2 \times 41.37\,\text{h} = 18,244.17\,\text{deg}^2\,\text{h}$. Let $N = 1$ be the number of detected events. Then, the FRB rate over the whole sky, $R$, in number of FRBs per hour is

\begin{equation}
    R = 4\pi\left(\frac{180}{\pi}\right)^2\frac{N}{E} = \frac{1\times 41253 \,\text{deg}^2}{18244.17\,{\rm deg}^2\,{\rm h}} = 2.26\,{\rm sky}^{-1}\,\text{h}^{-1},
\end{equation}

or $54.24\,{\rm sky}^{-1}\,{\rm day}^{-1}$. Given the small sample size, Poisson statistics was used to determine the uncertainty \citep{Gehrels1986}. The 95\% confidence interval is $1\text{-}302\,{\rm sky}^{-1}\,{\rm day}^{-1}$.

The reference fluence threshold was calculated by scaling the 3.92 Jy peak flux density of \frbname{} detected with a S/N of 6.84 to the fluence corresponding to the detection S/N threshold of 5. The fluence threshold of the search therefore is $3.92 \times (5/6.83) \times 20\,\text{ms} \approx 57$\,Jy\,ms.

\bibliography{sn-bibliography}

\section*{Data availability}
Access to raw voltage data of the SMART survey is at discretion of the MWA director and are subsequently accessed through the MWA ASVO portal. Dynamic spectra of the FRB at 20\,ms and 1\,ms time resolutions will be made available through the Transient Name Server\footnote{\url{https://www.wis-tns.org}}, and can also be requested from the Corresponding Author.

\section*{Code availability}
The BLINK FRB search pipeline, including the implementation of the STRIDE algorithm, is publicly available on GitHub: \url{https://github.com/PaCER-BLINK-Project}.

\section*{Acknowledgments}
This scientific work uses data obtained from Inyarrimanha Ilgari Bundara / the Murchison Radio-astronomy Observatory. We acknowledge the Wajarri Yamaji People as the Traditional Owners and native title holders of the Observatory site. Establishment of CSIRO's Murchison Radio-astronomy Observatory is an initiative of the Australian Government, with support from the Government of Western Australia and the Science and Industry Endowment Fund. Support for the operation of the MWA is provided by the Australian Government (NCRIS), under a contract to Curtin University administered by Astronomy Australia Limited. 

This scientific work uses data from Murriyang, CSIRO’s Parkes radio telescope, part of the Australia Telescope National Facility (https://ror.org/05qajvd42) which is funded by the Australian Government for operation as a National Facility managed by CSIRO.

This work was supported by resources provided by the Pawsey Supercomputing Research Centre’s Setonix Supercomputer, with funding from the Australian Government and the Government of Western Australia. The authors also acknowledge the Pawsey Centre for Extreme Scale Readiness (PaCER) for funding and support.

This work was supported by the Australian SKA Regional Centre (AusSRC), Australia’s portion of the international SKA Regional Centre Network (SRCNet), funded by the Australian Government through the Department of Industry, Science, and Resources (DISR; grant SKARC000001). AusSRC is an equal collaboration between CSIRO – Australia’s national science agency, Curtin University, the Pawsey Supercomputing Research Centre, and the University of Western Australia.

This work has also been co-funded and supported by the International Centre for Radio Astronomy Research (ICRAR).

\section*{Author contributions}
This work is part of C.D.P's Ph.D. studies and will be the third paper of his thesis. C.D.P. designed and developed the BLINK FRB search pipeline starting from the GPU imager initially designed and developed by M.S, and extending it with the STRIDE dedispersion algorithm. C.D.P. designed, implemented, and executed the SMART processing on the Setonix supercomputer, including data staging, job scheduling and submission, and candidates inspection, which led to the discovery of the FRB. C.D.P. and M.S. conducted post-processing analysis and interpretation, with contributions from all co-authors. CDP and MS wrote the first draft of the manuscript, produced all figures, and iterated subsequent versions. All authors reviewed and provided comments on the draft manuscript. M.S. is the main supervisor and the author of the initial proposal for the Ph.D. project. M.S. significantly contributed to the development of the various steps of the BLINK pipeline with a particular focus on fast imaging, source localisation, and RFI identification. C.W.J. supported the post-processing with statistical analyses, aided interpretation, and validated methods adopted by C.D.P. and M.S. C.W.J. contributed his knowledge and expertise in FRBs during regular meetings. N.D.R.B. carried out follow-up observations with Murriyang at Parkes and contributed his expertise with the SMART dataset. R.W. contributed his expertise with the MWA and his knowledge of the theory of interferometry, both during the development of the pipeline and the SMART processing. C.P.L., B.W.M., and C.M.T. confirmed the FRB in beamformed data and helped with tools used typically in pulsar astronomy. M.S., C.W.J., N.D.R.B., R.W., D.C.P., and C.H. are part of the Ph.D. supervisory panel and contributed their time and effort throughout this multi-year project. N.D.R.B., C.P.L., B.W.M., S.E.T., G.S., M.X., and S.J.M. are members of the SMART collaboration who curated the SMART dataset, and provided useful comments on the manuscript. C.S. and M.E. are Pawsey technical staff members who actively facilitated the PB-sized data retrieval and staging on the /scratch filesystem over the course of months-long processing.

\section*{Competing interests}
The authors declare that they have no competing financial interests.

\newpage
\section*{Extended data}
\renewcommand{\figurename}{Extended Data Fig.}
\setcounter{figure}{0}

\begin{figure}[h!]
    \centering
    \includegraphics[width=\linewidth]{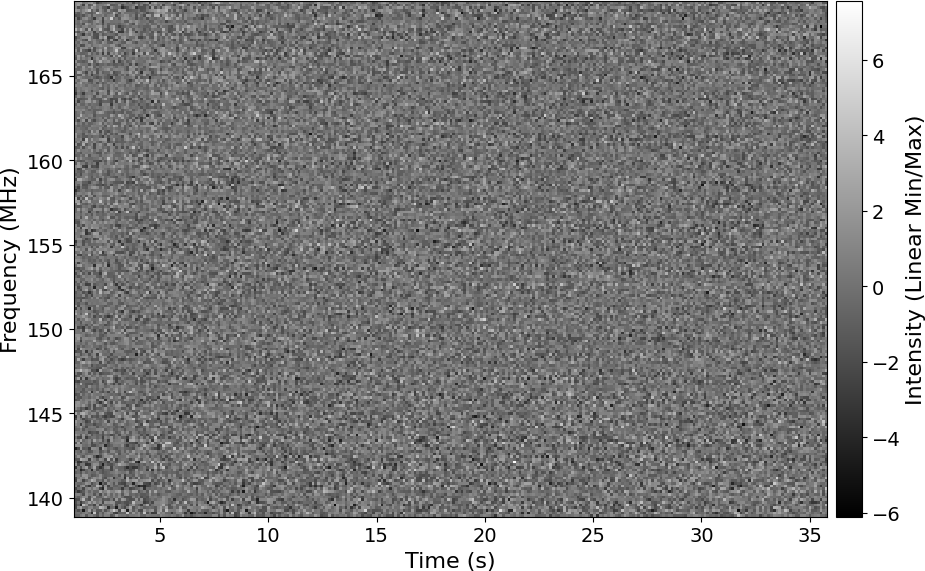}
    \caption{\textbf{Dispersive sweep of \frbname{}}. Depicted in the figure is the dynamic spectrum before dedispersion in the 128\,ms time resolution that maximises the S/N of the dedispersed pulse. A weak, dispersed FRB pulse spans the entire band, closely following the expected $\nu^{-2}$ dispersion sweep. In our best-fit model of the $\nu^{-\gamma}$ dependence we find $\gamma=2.005^{0.010}_{-0.002}$ at the 68\% confidence level, which is consistent with the $\nu^{-2}$ dispersive delay expected from propagation through cold ionized plasma.}
    \label{fig:dispersion_sweep}
\end{figure}

\begin{figure}[h!]
    \centering
    \includegraphics[width=\linewidth]{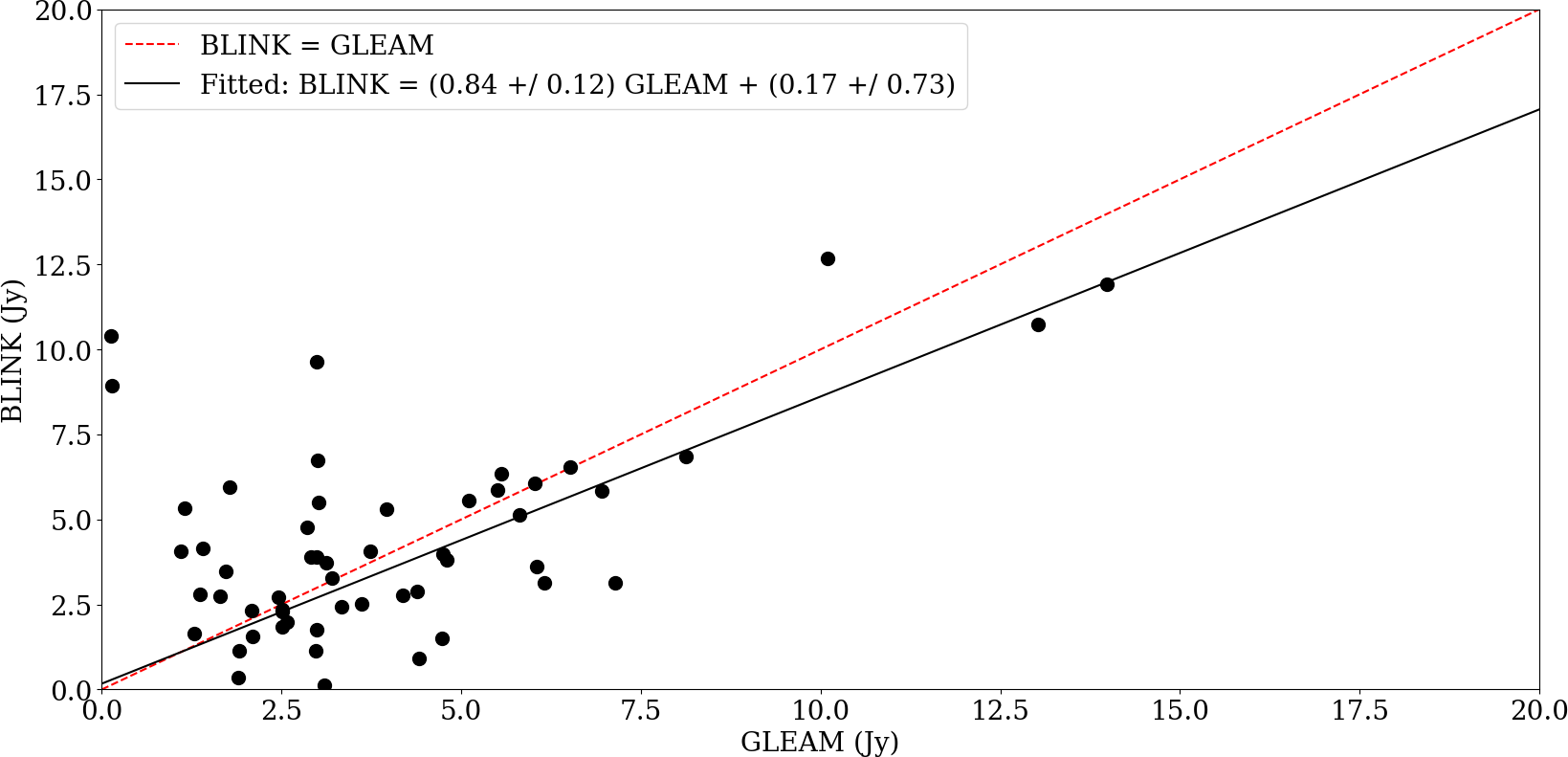}
    \caption{\textbf{Flux density calibration.} The flux density in the calibrated BLINK dirty images, labelled as BLINK, is compared to the integrated flux density of GLEAM sources within a 13\,arcmin matching radius, labelled as GLEAM. The red dashed line is the one-to-one line (BLINK = GLEAM), and the black line is a fit to sources with an integrated GLEAM flux above 3\,Jy. The plot demonstrates that the flux densities measured in the BLINK dirty images, after the corrections described in the text, are correctly calibrated and match the GLEAM flux scale. The distribution of these differences yields a relative uncertainty of approximately 50\%, primarily driven by sidelobe and classical confusion noise in the dirty image. For comparison, flux densities derived from the CLEANed WSCLEAN image lead to a relative uncertainty of around 20\%. Error bars ($\pm$2\,Jy) are omitted for clarity.}
    \label{fig:blink_vs_gleam_flux}
\end{figure}

\begin{figure}
    \centering
    \includegraphics[width=\linewidth]{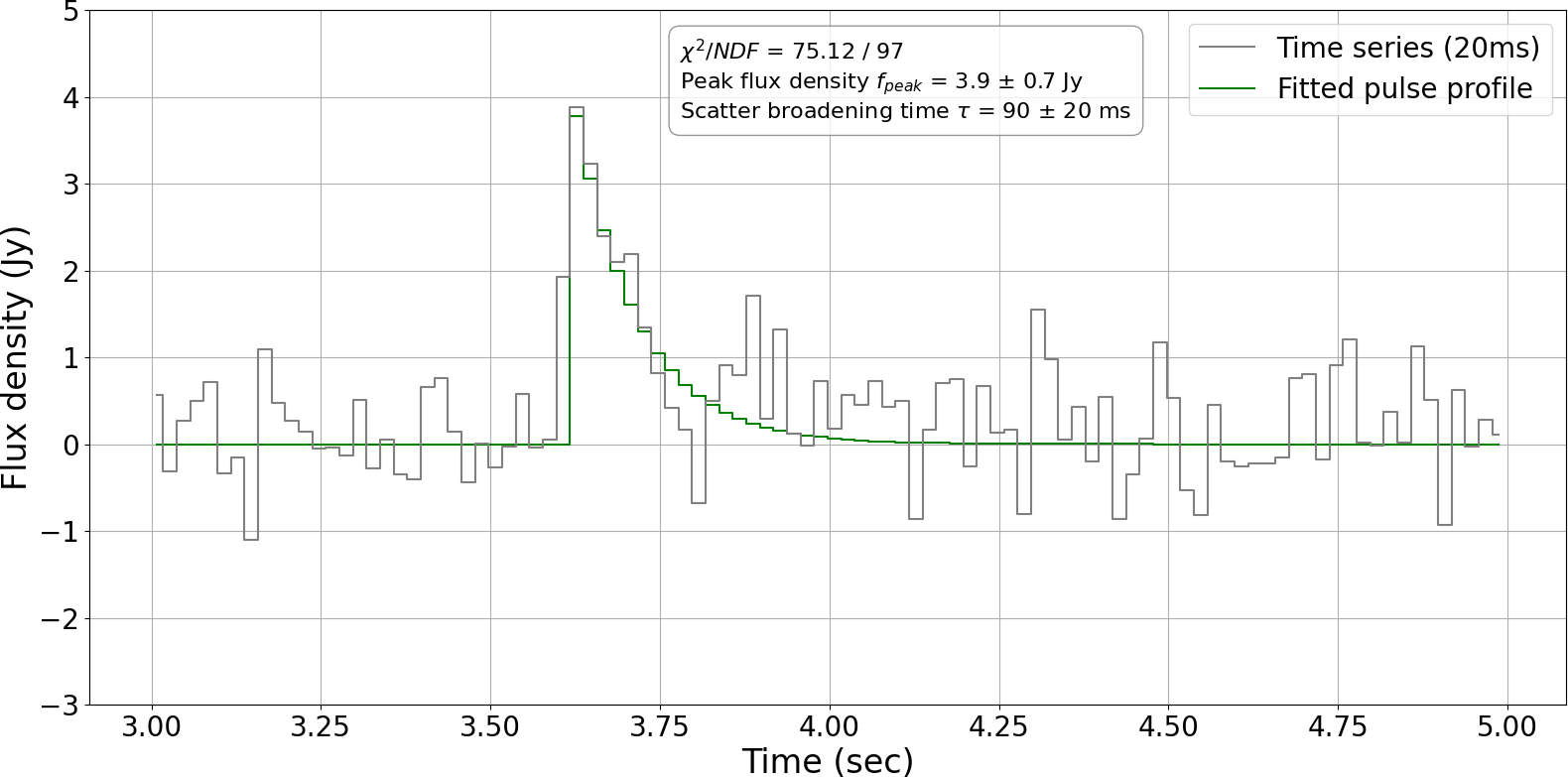}
    \caption{\textbf{Scatter broadening profile of \frbname{}.} Flux density as a function of time at a 20\,ms resolution (black curve), overlaid with a best-fit Gaussian convolved with an exponential tail (green curve). The scatter broadening time derived from this fit is $\tau =90\pm20$\,ms. The uncertainty reflects the time resolution used. Hence, the intrinsic pulse width is less than $20$\,ms. Error bars ($\approx$0.7\,Jy) are omitted for clarity. Fits to the 5\,ms and 10\,ms time series yield consistent results and are omitted for brevity.}
    \label{fig:scatter_broadening_time_fit}
\end{figure}

\end{document}